\documentclass{article}
\usepackage{epsfig}
\usepackage{graphicx}
\begin{document}

%\usepackage{graphicx}% Include figure files
%\usepackage{dcolumn}% Align table columns on decimal point
%\usepackage{bm}% bold math
%\usepackage{color}
%\usepackage{ulem}
%\begin{document}
\title{{\bf Geometric Phase: a Diagnostic Tool for Entanglement}\\
S. N. Sandhya$^{1}$ \footnote{email:sns@iitk.ac.in} and Subhashish Banerjee $^{2}$ \footnote{email:subhashish@iitj.ac.in} \\ 
%\address{
$^{1}$ Department of Physics, IIT Kanpur, Kanpur 208016, India \\
$^{2}$ IIT Rajasthan, Jodhpur 342011, India}

%\pacs{03.65.Vf}{First pacs description}
%\pacs{03.65.Ud}{Second pacs description}
%\pacs{03.65.Wj}{Third pacs description}

\maketitle
\abstract
Using a kinematic approach we show that the non-adiabatic, non-cyclic, geometric phase corresponding to the radiation emitted by a three level cascade system provides a sensitive diagnostic tool for determining the entanglement properties of the two modes of radiation. The nonunitary, noncyclic path in the state space  may be realized 
through the same control parameters which control the purity/mixedness and entanglement. We show analytically that the geometric phase is related to concurrence in certain region of the parameter space. We further show that the rate of change of the geometric phase reveals its 
resilience  to fluctuations only for pure Bell type states. Lastly, the derivative of the geometric phase carries information on both purity/mixedness and entanglement/separability.

%\pacs{32.80 Qk, 42.50 Gy,}
%\maketitle
\section{Introduction}
Geometric phase \cite{liter} and topological phases have provided impetus\cite{sjoquist_physics} to quantum information processing. Of the two geometric phase is easier for implementation.
Both these approaches  may be designed to  realize fault tolerance \cite{knill}  in addition to  resilience 
to decoherence \cite{wu, oreskov}. Simple quantum gates using geometric phase have been demonstrated experimentally in the 
nuclear magnetic resonance set up \cite{jones}. Geometric phase  has been studied in its entirety encompassing 
various aspects of evolution including adiabatic \cite{adiabatic}, cyclic, unitary  and subsequently the generalization to non-adiabatic \cite{non-adiabatic},  non-cyclic 
\cite{noncyclic,nonunitary}, and non-unitary \cite{nonunitary,Tong} evolution using either a dynamic or a kinematic or approach 
\cite{kinematic, Tong}. Non-unitary evolution in the context of open systems has also been reported \cite{refB}. These studies have been carried out both for  pure and  mixed states \cite{mixed, bs08}. On the other hand an operational 
definition of the geometric phase has been given  by Sj\"oqvist et al \cite{operational}. There have been further studies on the relationship of 
the geometric phase of an entangled system and its sub-system \cite{entgl}. Recently, a proposal for studying the 
Pancharatnam phase  and its relation to non-local quantum correlations has been suggested\cite{sam_supurna} in the two-photon interferometric set up.

Holonomic quantum computation and entanglement require the evolution of  qubits in  parameter space by the control of parameters which are physically feasible.  Both geometric phase and entanglement  are important tools in  quantum information processing. Atom photon interaction provides common ground for exploring the relationship between  geometric phase and entanglement with the added advantage of the existence of  control parameters for manipulating  the two photon states. In this letter we study the non-adiabatic, non-cyclic, non-unitary evolution of the geometric phase of the two-photon state corresponding to the two modes emitted by a three level cascade system  interacting with two driving fields \cite{sns_vr}. The two photon state is in general a mixed state, however it may also be prepared in a pure state by proper control. We consider the evolution in the state space by varying the control parameters namely,  the driving field strength and detuning. 
\section{The Model}
The scheme considered here  corresponds to a three level cascade system interacting with two coherent fields which address the only two allowed dipole transitions $|i\rangle \leftrightarrow |i+1\rangle, i=1,2$ with energy separation given by $\omega_i$. Two counter propagating ( Doppler free geometry) driving fields of nearly equal frequencies $\omega_{L1}$ and $\omega_{L2}$ and respective strengths $\Omega_1$ and $\Omega_2$ are resonant with these two transitions. The decay constants of the energy levels  $| 3 \rangle $ and $| 2\rangle $ are indicated by $ \Gamma_3$ and $\Gamma_2$ respectively. The parameters $\Delta_1,~ \Delta_2 $  refer to the detunings of the driving fields. This scheme may be realized, for example, in $^{87}Rb$ vapor with the corresponding energy levels $5S_{1/2}, 5P_{3/2}$ and $5D_{5/2}$ and has been used by Banacloche {\em et al.} \cite{banacloche} for studying  electro-magnetically induced transparency. The parameters used for the numerical computation correspond to this scheme which we label as scheme I.
\begin{figure}
\centering
\includegraphics[height=5cm]{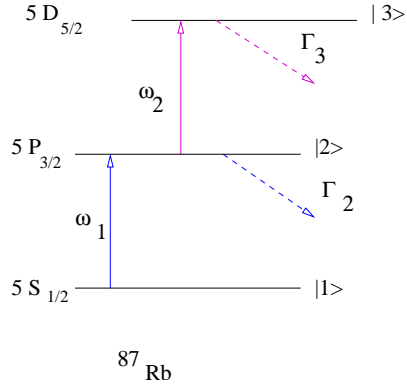}
\caption{(Color online) Three level cascade system corresponding to the $^{87}Rb$ atoms driven by two fields
$\omega_1$ and $\omega_2$.}
\label{fig:1}
\end{figure}

The driven three level atomic system emits radiation corresponding to the dipole transitions $|i+1\rangle \rightarrow | i\rangle$.The resulting field is a two photon state in the Fock space
and its  density matrix is of rank three which follows from Schmidt decomposition. This class of two photon states is in a smaller sub-space and requires just $(N+1)^2-1$ variables rather than $4^N$ variables to completely determine the $N=2$ state. This reduction simplifies the analysis enormously.

The available control parameters for tailoring  the two photon state are the driving field strengths $\Omega_i, ~i=1,2$ and the detunings 
$\Delta_i,~i=1,2.$ corresponding to the two transitions. Let us denote the two photon detuning by  $\Delta=\Delta_1+\Delta_2$. At two photon resonance $\Delta=0$. Some of the non-classical properties of this two photon field have been reported in the literature \cite{clauser, sns_vr}. In particular,  a detailed numerical study of the entanglement distribution as a function of the these control parameters has been reported in \cite{sns_vr}.      
In order to establish the dependence of entanglement and purity on these control parameters we make a small change in the scheme used above, namely, we replace the metastable state ($\Gamma_3=1 MHz$) by an infinitely long lived state i.e, $\Gamma_3=0$  which we call scheme II. In this case, the two photon decoherence given by $(\Gamma_3+\Gamma_1)/2$ is zero which results in a two photon system   with no decoherence. This simplifies the analysis and (i) the dependence of the entanglement on the control parameters and (ii) the relation between geometric phase and entanglement are more transparent. The effect of  decoherence is included in the real system.
%We subsequently justify the use of scheme II by showing that the physics is qualitatively the same.

The two photon density matrix for scheme I (Fig.1) may be determined
 using the quantum tomographic method described in \cite{sns_vr}. We briefly recapitulate the procedure here. The atom+field pure state, at a time $t$, is written as
\begin{equation}
|\Psi (t) \rangle = \sum_{i,n} \alpha_{i}^{n} (t) |i ; n \rangle; i=1, 3; n=0,1, 2, 3;
\end{equation}
where $i$ denotes the atomic index while $n$ denotes the photon mode index. To be more specific, in the binary notation $n = n_2~n_1$, where $n_i=0,1$. The two photon state may now be obtained by taking the partial trace over the atomic indices. It has been further shown \cite{sns_vr}, that the  two photon density matrix at a time t is equivalent to the atomic density matrix at a retarded time $t-r/c$ \cite{bibitem}. Utillizing this equivalence we write 
 the two-photon normalized
 density matrix  in the basis $\{ |0\rangle, |1\rangle, |3\rangle\} \equiv \{ |00\rangle, |01\rangle, |11\rangle\}$ as
\begin{eqnarray}
\rho^{n}(t)&=&\frac{1}{\cal N}\left (
\begin{array}{ccc}
\sum_i |\alpha^0_i|^2 & \sum_i \alpha^0_i\alpha^{1^*}_i   & \sum_i \alpha^0_i\alpha^{3^*}_i \\
... &\sum_i |\alpha^1_i|^2   & \sum_i \alpha^1_i\alpha^{3^*}_i \\
... &...  &\sum_i |\alpha^3_i|^2
\end{array}
\right ) \nonumber \\
&\equiv& \rho^A(t-r/c)
\end{eqnarray}
Where ${\cal N}$ is the normalization factor. The solution for the complex coefficient, $\alpha^{n}_i(t)$ may be obtained by solving the Liouville equation., 
\begin{equation}
i \hbar \dot \rho^A= [ H, \rho^A] -\frac{i\hbar}{2}\{\Gamma,\rho^A\}
\end{equation}
where $\Gamma$ is the relaxation matrix and $H$ is the Hamiltonian  given by 
\begin{eqnarray}
 H &=& \frac{\hbar}{2}  ( \omega_1 \sigma^z_{1} + \omega_2 \sigma^z_{2} )+
 \hbar \Omega_1 (e^{-i \omega_1 t} \sigma^{+}_{1} + h.c.) +\nonumber  \\
& & \hbar \Omega_2 (e^{-i \omega_2 t} \sigma^{+}_{2} + h.c.)
\end{eqnarray}
in the interaction picture and in the rotating wave approximation. The Rabi frequency is given by   $\Omega_i= -\frac{1}{\hbar}\vec {\mu}_{i i+1} \cdot \vec{E}_i;~ i=1,2$, corresponding to the two driving field strengths.   $\sigma^{+}_{i}= |i+1 \rangle ~\langle i|; ~\sigma^{-}_{i}= |i \rangle ~\langle i+1|; ~ i = 1, 2 $ and $\sigma^z_{i} = |i+1 \rangle ~\langle i+1|- |i \rangle ~\langle i|$ are the  atomic transition operators.
\begin{figure}
\centering
\includegraphics[width=5cm]{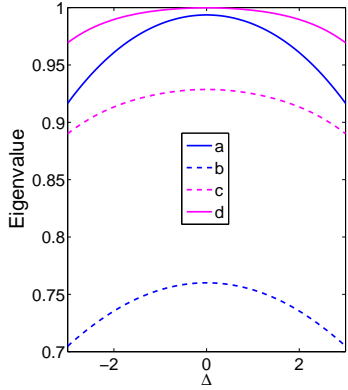}
\caption{(Color online) Variation of the eigenvalues  with $\Delta$. The solid lines correspond to scheme II while the dashed lines correspond to scheme I. Parameter values are a) and b)    $\Omega_1=6, ~\Omega_2=6$ and c) and d) $\Omega_1=3,\Omega_2=6$.}
\label{fig:3}
\end{figure}
\begin{figure}
\centering
\includegraphics[width=5cm]{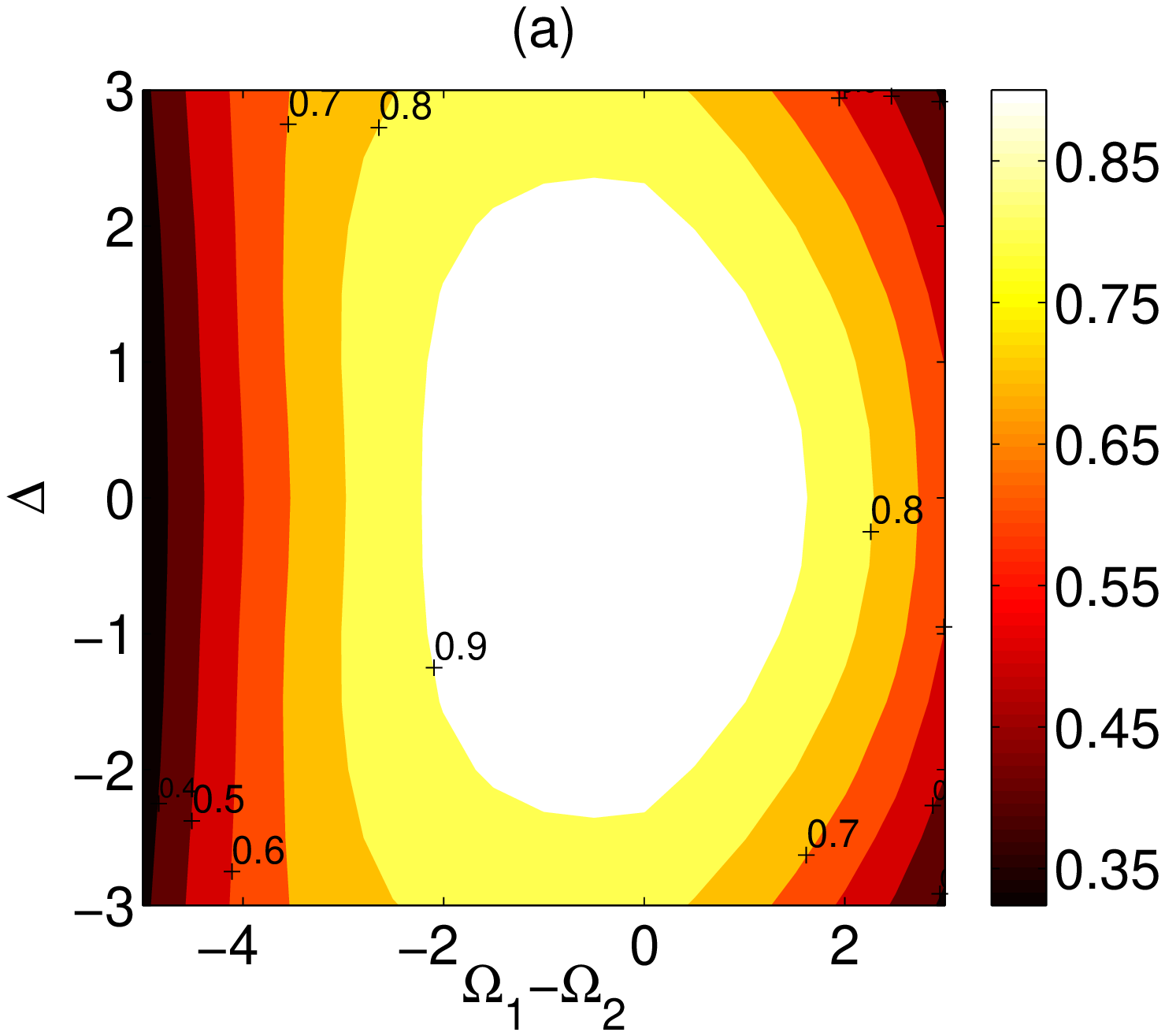}\\
\includegraphics[width=5cm]{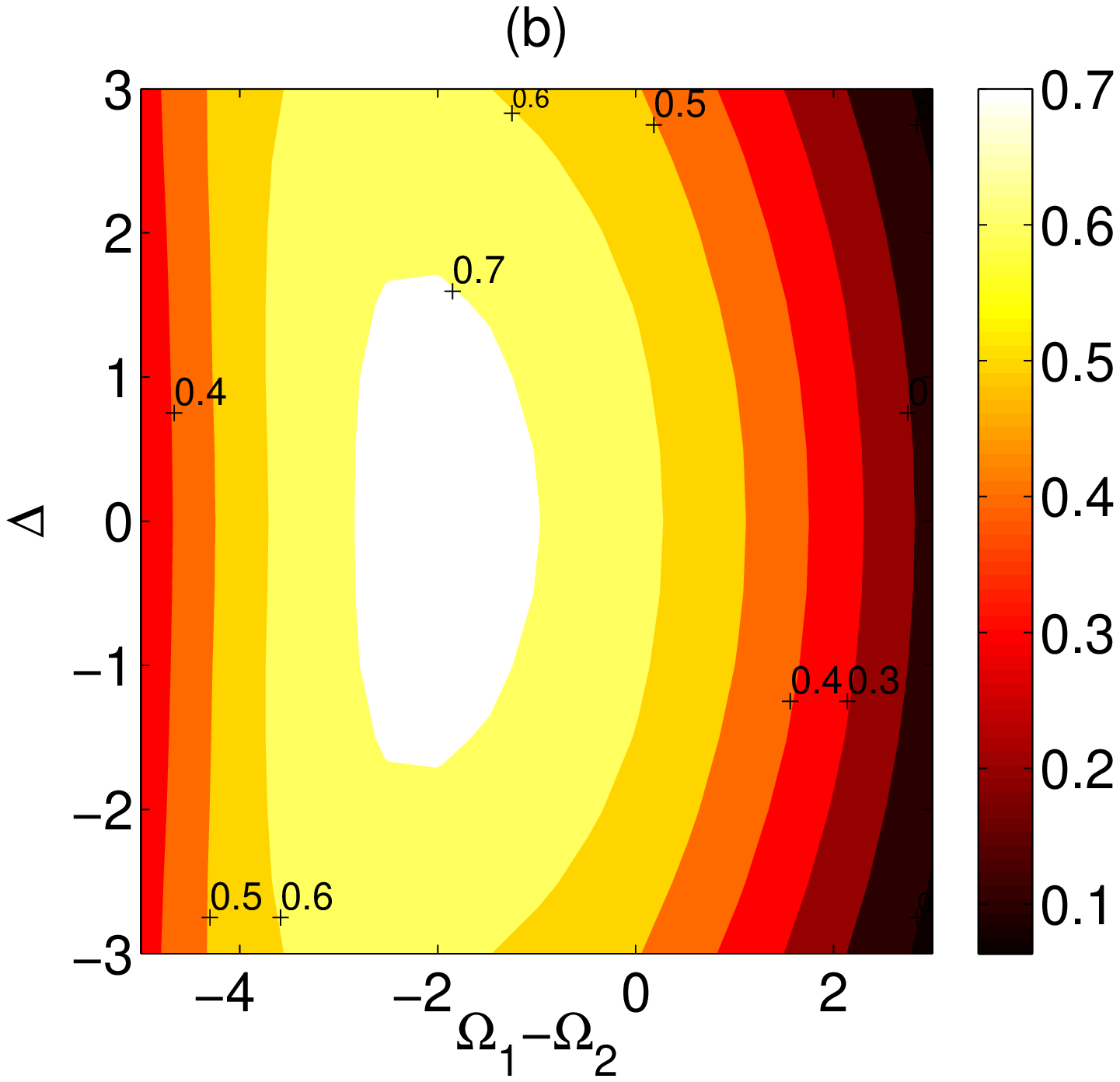}\\
\caption{(Color online) Variation of concurrence in the parameter space $(\Delta,X)$. (a) Ideal system (scheme II) and (b) Real system (scheme I) with $\Omega_2=6$.}
\label{fig:3}
\end{figure}
\section{Geometric phase}
The Pancharatnam relative phase for a pure state  is defined as
\begin{equation}
\alpha (t)= Arg(\langle \Psi(0)| \Psi(t)\rangle).\label{panch}
\end{equation}
The two photon state under consideration, on the other hand is in general a mixed state. We follow here the kinematic approach of Tong {\em et al.} \cite{Tong}  for determining the geometric phase for non-cyclic, non-unitary, non-adiabatic evolution of mixed states.  We briefly recapitulate their procedure  here: A general mixed state density matrix $\rho(t)$ is subject to purification by the introduction of an ancilla as
\begin{equation}
|\Psi(t)\rangle =\sum_k \sqrt{\lambda(k)} |\phi_k(t)\rangle \otimes |a_k\rangle ; t \in [0,\tau].
\end{equation}
The Pancharatnam relative phase given in Eq. (\ref{panch}) reduces to the geometric phase when the parallel transport condition is satisfied. Tong {\em et al.} \cite{Tong} impose  the parallel transport condition for the mixed state to be $\langle \phi_k(t) | d/dt |\phi_k(t)\rangle =0, k=1...N$ corresponding to the $N$ eigenstates. In the scheme under consideration $N=1, 2, 3$. The geometric phase for the mixed state, $\rho^{n}(t)$, satisfying the parallel transport conditions assumes the form
\begin{eqnarray}
\gamma_g(\tau)&=&Arg \big[\sum_k\sqrt{\lambda_k(\tau)\lambda_k(0)}\langle \phi_k(0)|\phi_k(\tau\rangle  \times \nonumber \\
& &e^{-\int_0^{\tau} \langle \phi_k(t')|
\dot \phi_k(t')\rangle dt'}\big] \label{GP}
\end{eqnarray}
where $\lambda_k(\tau)$ are the eigenvalues and $\phi_k(\tau)$ are the corresponding eigenvectors. Here the evolution is considered over the parameter space consisting of $\{\Delta, X \}$ defined in the next section.

\section{Relation between Geometric phase and Entanglement}
In order to get a physical understanding of the dependence of the two photon state on the free parameters, we consider first, the  density matrix of the ideal system (scheme II) near two photon resonance which has the simple form
\begin{eqnarray}
\rho&=&\left (
\begin{array}{ccc}
{\rm S}^2& \bar \Delta {\rm C} {\rm S}^2 & -{\rm S} {\rm C} (1- i \gamma_{21} \bar \Delta )\\
... &0 & -\bar \Delta \rm{C}^2 {\rm S} \\
... &...  &{\rm C}^2
\end{array}
\right ) 
\end{eqnarray}
where $ {\rm S}={\rm Sin}(X) ~{\rm and}~  {\rm C}={\rm Cos}(X)$. We have defined the dimensionless parameters as  $X={\rm Tan}^{-1}(\Omega_1/\Omega_2), \bar  \Delta=\Delta/ \sqrt{\Omega_1^2+\Omega_2^2}$ and $\gamma_{21}=\Gamma_{2}/2\sqrt{\Omega_1^2+\Omega_2^2}$. The density matrix is first order in the parameter $\bar \Delta$. The reason for choosing the state around this point will become clear shortly.
Consider the spectral resolution of the density matrix 
\begin{equation}
\rho=\sum_i \lambda_i |\psi_i\rangle \langle \psi_i \vert;~ i=1,3
\end{equation}
where $\lambda_i$ are the eigenvalues and $|\psi_i \rangle$ are the corresponding eigenstates of the density matrix. 
The steady state eigenvalues at $\bar \Delta=0$   is given by
$\{0,0,1\}$,  which means it is a pure state.  While in the real system (scheme I) the largest eigenvalue is always $< 1$. However, for any  given $\Omega_1,\Omega_2$,  purity is still maximum for $\bar \Delta=0$ with the largest eigenvalue being maximum at this point. A comparison of the eigenvalues for the ideal and real system  is illustrated  in Fig.2. The role of the metastable state thus, is to  introduce a  admixture, due to decoherence, to an otherwise pure state.  Since the atomic decay constants are fixed in a real system, the only remaining free parameter for controlling purity is   the two photon detuning $\bar \Delta$. The state  around the point 
$(\bar \Delta=0, X)$ may be approximated to a pure state when there is no decoherence. This is the reason for choosing to study the evolution near two photon resonance. In the ideal case ( scheme II) the pure state at $\bar \Delta=0$ is given by $|\Psi^{pure}\rangle = -{\rm Sin}(X) |00 \rangle + {\rm Cos}(X) |11\rangle; ~ {\rm Sin}(X)= \Omega_1/ \sqrt{\Omega_1^2+\Omega_2^2}$.  
 The concurrence for this pure state is  ${\cal C}=2 {\rm Sin}(X) {\rm Cos}(X)$ and is maximum for $X=\pi/4$ or 
 $\Omega_1=\Omega_2$ which we call the bell state regime since this corresponds to the Bell state
 $ (-|00\rangle +|11\rangle)/ \sqrt{2}$.  On the other hand concurrence is very small when $\Omega_1<< \Omega_2$. The state is separable in this region and we call this the separable state regime.  A non-perturbative evaluation  of the distribution of concurrence in the parameter space  for the system without decoherence is illustrated  in Fig.3a. It is clear that ${\cal C}$ is maximum in the region where the ratio of the field strengths is approximately one while the state becomes less entangled as the ratio between $\Omega_1$ and $\Omega_2$ differs from one. Thus the parameter $X$ which is a function of $\Omega_1, \Omega_2$ is the control parameter for entanglement.  Having established that $\bar \Delta$ and $X$ are the control parameters for purity and entanglement we now show that the geometric phase is related to entanglement.
 
We evaluate the  geometric phase in the space spanned by the parameters $\{\bar \Delta, X \}$ in the neighborhood of $\chi_0=(0,X)$ where $\lambda_i(0,X)=0,~\forall X, ~i=1,2 $
 and $\lambda_3(0,X) =1,~\forall X$. In other words we need to evaluate the geometric phase of the pure state at $\chi=(\delta,X+ dX)$ which is essentially the Pancharatnam's phase given by
%\begin{eqnarray}
%\gamma_g(\Delta, X)&=&Arg[\sqrt{\lambda_3(\Delta, X)}\langle \phi_3(0, X_0)|\phi_3(\Delta, X) \rangle  \times \nonumber \\
%& &e^{-\int_0^{X_i} \langle \phi_3(X_i')|
%\dot \phi_3(X_i')\rangle dt'}] \label{GP}
%\end{eqnarray}
\begin{equation}
\gamma_g(\chi)=Arg[\langle \psi_3(\chi_0)|\psi_3(\chi) \rangle ]
\end{equation}
here $\psi_3$ corresponds to the pure eigenstate and $\chi= \chi_0 + d\chi=(\delta, X+dX)$ is any point chosen infinitesimally close to $\chi_0$. Taylor expanding $\gamma_g$ near $\chi_0$ and retaining up to second order in $\delta$ and $dX$ we obtain
\begin{eqnarray}
\gamma_g(\chi)&=&Arg[1-\frac{dX^2}{2}  + \beta ~\delta^2- i  \gamma_{21} {\rm Cos} (X)^2 \delta -\nonumber \\
& &\frac{i  \gamma_{21} {\rm Cos}(X) {\rm Sin}(X) ~\delta~dX}{2}]\nonumber \\
&= &Arctan[\frac{- i  \gamma_{21} {\rm Cos} (X)^2 \delta- i  \gamma_{21} {\cal C} ~\delta~dX/ 4}{ 1-\frac{dX^2}{2}  + \beta~ \delta^2} ]\nonumber \\
\end{eqnarray}
where 
\begin{eqnarray}
\beta&=&(-(1/8) {\rm Cos}(X)^4 (4 + 16 \gamma_{21}^2 - (5 + 8 \gamma_{21}^2) {\rm Cos}(2 X) +\nonumber \\
&& {\rm Cos}(4 X))+{\cal C}^2((1 + 8 \gamma_{21}^2) {\rm Cos}(2 X) + {\rm Cos}(4 X))/16 \nonumber
\end{eqnarray}
 and
 ${\cal C}$ is the concurrence at $(0,X)$. We have thus, shown that the geometric phase $\gamma_g(\chi)$ at $\chi$ near $\chi_0$ depends on the concurrence ${\cal C}$ at $\chi_0$ for the ideal system (scheme II).
  
  We next evaluate the derivative $\gamma_g'$ of the geometric phase  with respect to the two photon detuning $\bar \Delta$ which is shown in Fig.4.  To be more specific, we evaluate $\gamma_g'$ in the neighborhood of $(\bar \Delta=0, X)$. We have already mentioned that  when $X=0$, concurrence  ${\cal C}=0$ and we have a separable state. Let us examine the behavior of geometric phase around this point.  The rate of change of the geometric phase in  the neighborhood of $(\bar \Delta=0,X=0)$
varies  rapidly for small variations   (Fig.4 (a))  $\delta$ along the $y$ axis and seems to be unchanged for small variations $dX$ along the  $x$ axis. On the other hand when $X=\pi /4$ concurrence ${\cal C}=1$, which is a Bell state. Here, the rate of variation of the geometric phase is relatively very slow  for small perturbation in both $\delta$ and $dX$. This is illustrated in Fig.4b. While in the case of ${\cal C}=1,$ the total change of the slope, $\gamma_g'$, is from -1.4 to -2.7 (Fig.4b), in the case of ${\cal C}=0$ it changes from -3.0 to -14.0 (Fig.4a) for the same range of values of $\delta$ and $dX$. Thus the geometric phase seems to change very slowly in the vicinity of maximum entanglement.  Faster the sweep of the geometric phase or larger  the rate of change of geometric phase implies weaker the entanglement. In other words,  stability of the geometric phase in any region of the parameter space seems to indicate the state has maximum entanglement in this region.
In the general mixed case, however, it is difficult to  show analytically the relation between entanglement and geometric phase. We therefore illustrate the relation through numerical simulations.  
\begin{figure}
\centering
\includegraphics[height=4.2cm]{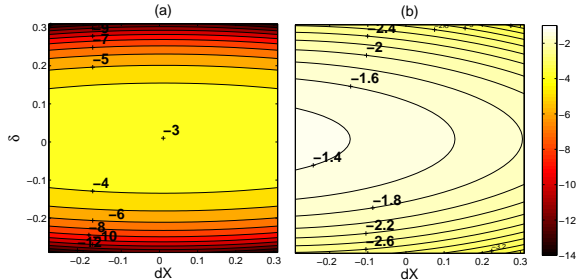}
\caption{(Color online) Variation  of  $d~\gamma_g/d~\bar \Delta$  in the neighborhood of (a)   $\bar \Delta=0$ and $X=0$, (separable state) and
(b) $\bar \Delta=0$ and $X=\pi/4$ (Bell state).}
\label{fig:4}
\end{figure}
%show the derivative of gp 
\section{Numerical Results:}
We present below the numerical results for the real (scheme I) system shown in Fig.1.
%\begin{figure}
%\centering
%\includegraphics[width=10cm]{fig2a.eps} 
%\includegraphics[width=6cm]{fig2b.eps} \\
%\includegraphics[width=6cm]{fig2c.eps} 
%\includegraphics[width=6cm]{fig2d.eps} 
%\caption{Variation of the geometric phase with the  rescaled detuning parameter $\delta_1=(\Delta_1+10.0)/20.0$. The parameter values are: a) $\Omega_1=\Omega_2=6.0, \Delta_2=0$, b) $\Omega_1=3.0, \Omega_2=6.0, \Delta_2=0$,
%%c) $\Omega_1=6.0, \Omega_2=3.0, \Delta_2=0$, d)$ \Omega_{1,2}=6.0, \Delta_2=3.0$ and e) $\Omega_{1,2}=6.0, \Delta_2=6.0.
%}
%\label{fig:2}
%\end{figure}
\begin{figure}
\centering
\includegraphics[width=6.5cm]{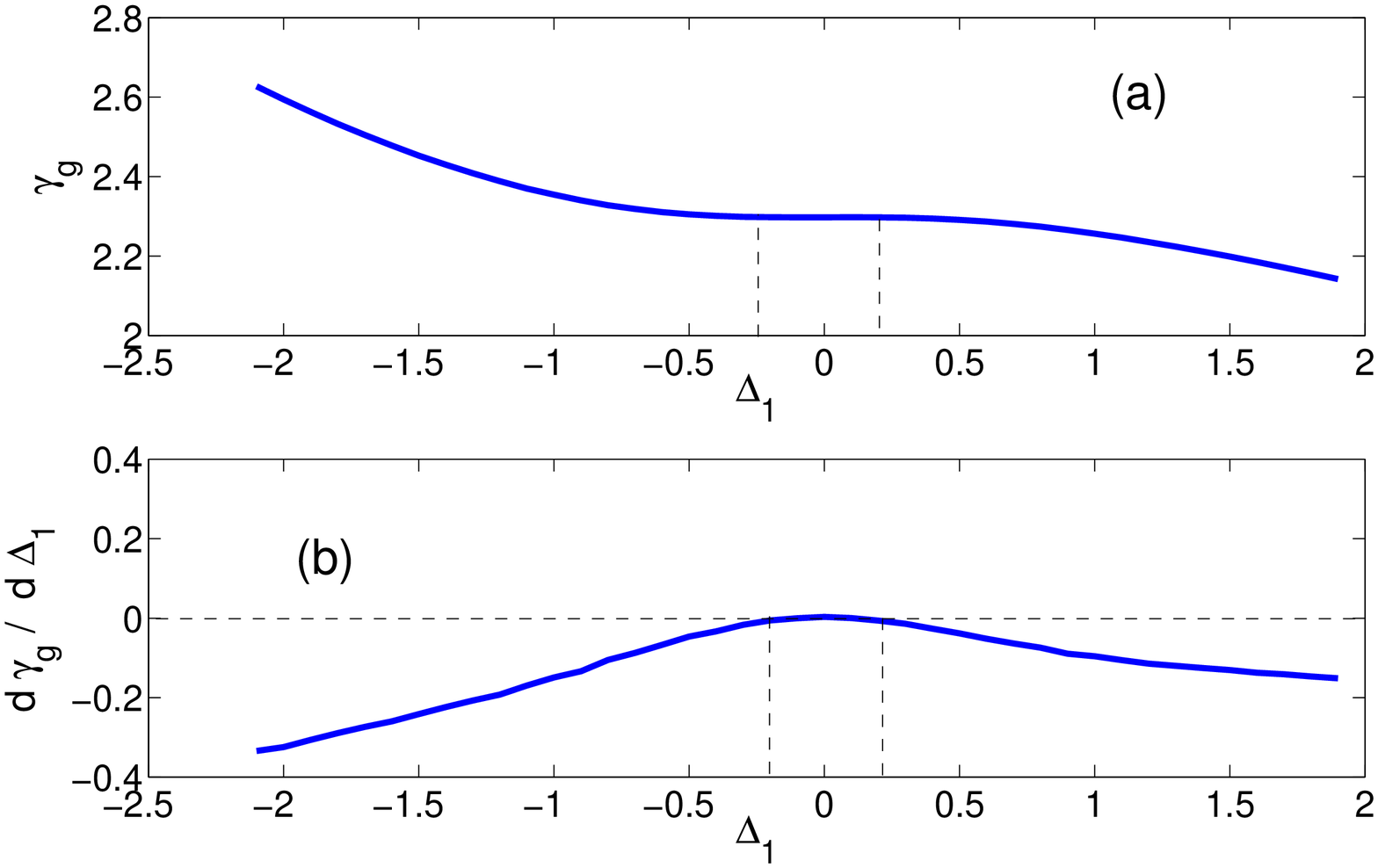} 
\includegraphics[width=6.5cm]{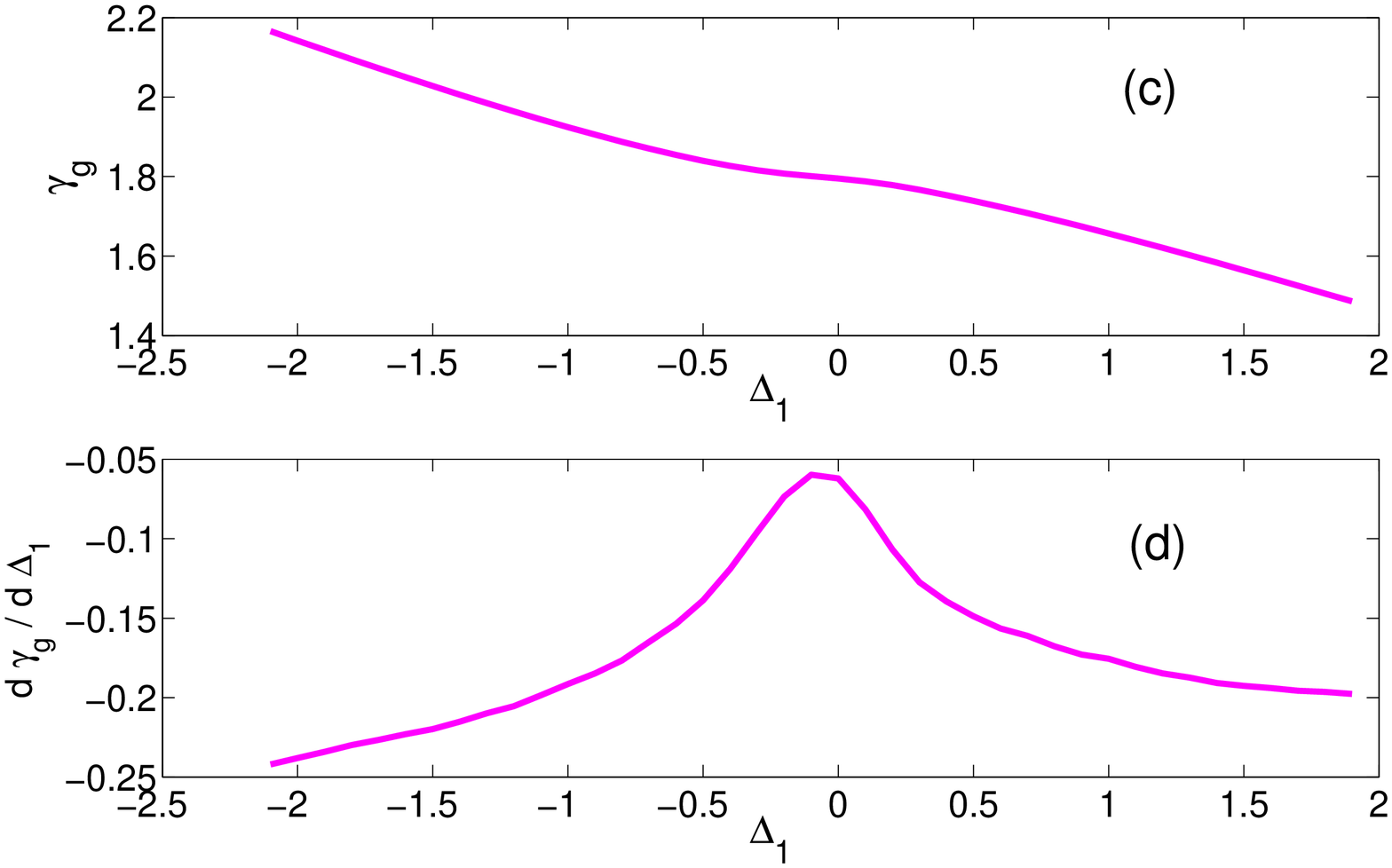} \\
\includegraphics[width=6.5cm]{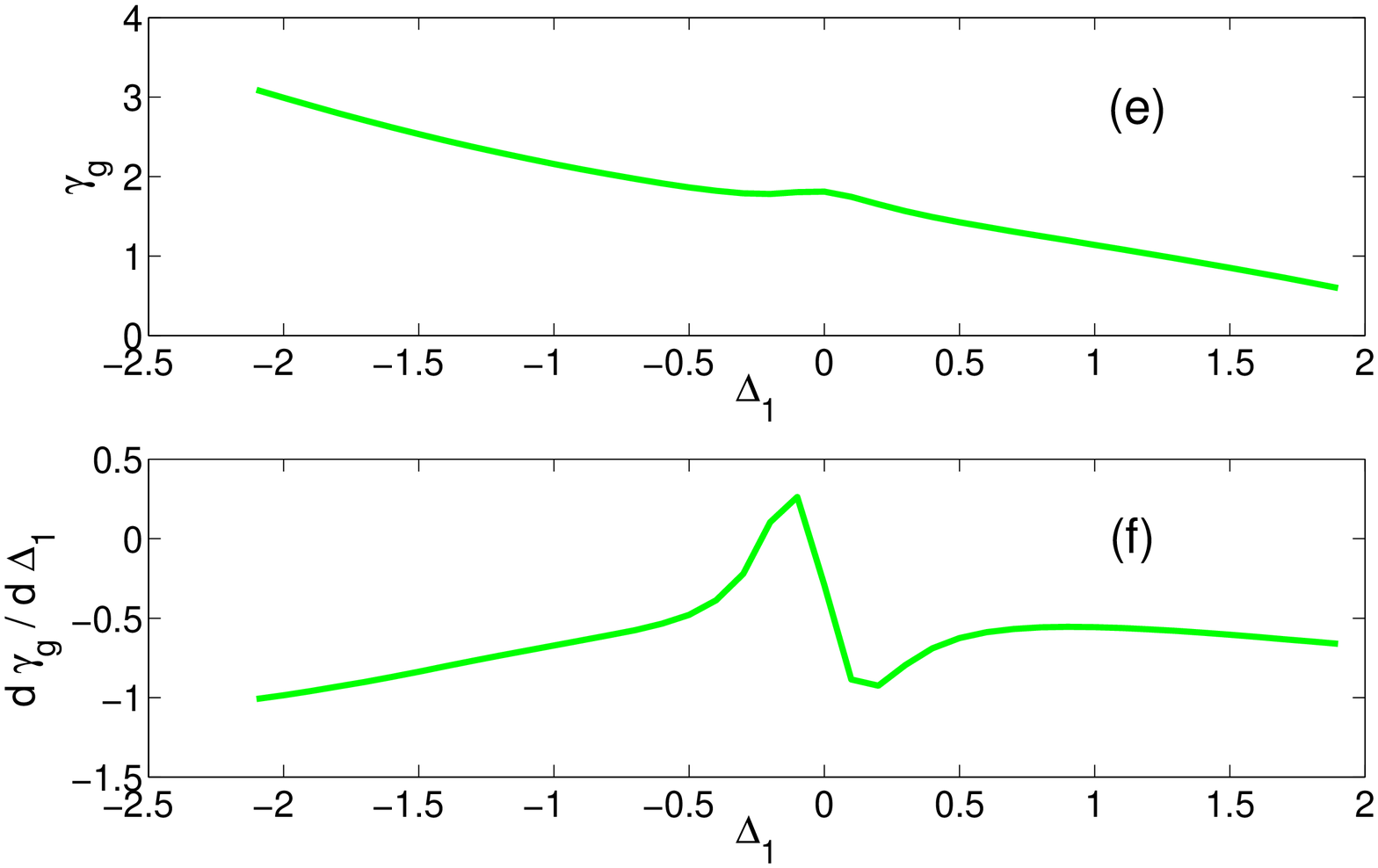} 
\includegraphics[width=6.5cm]{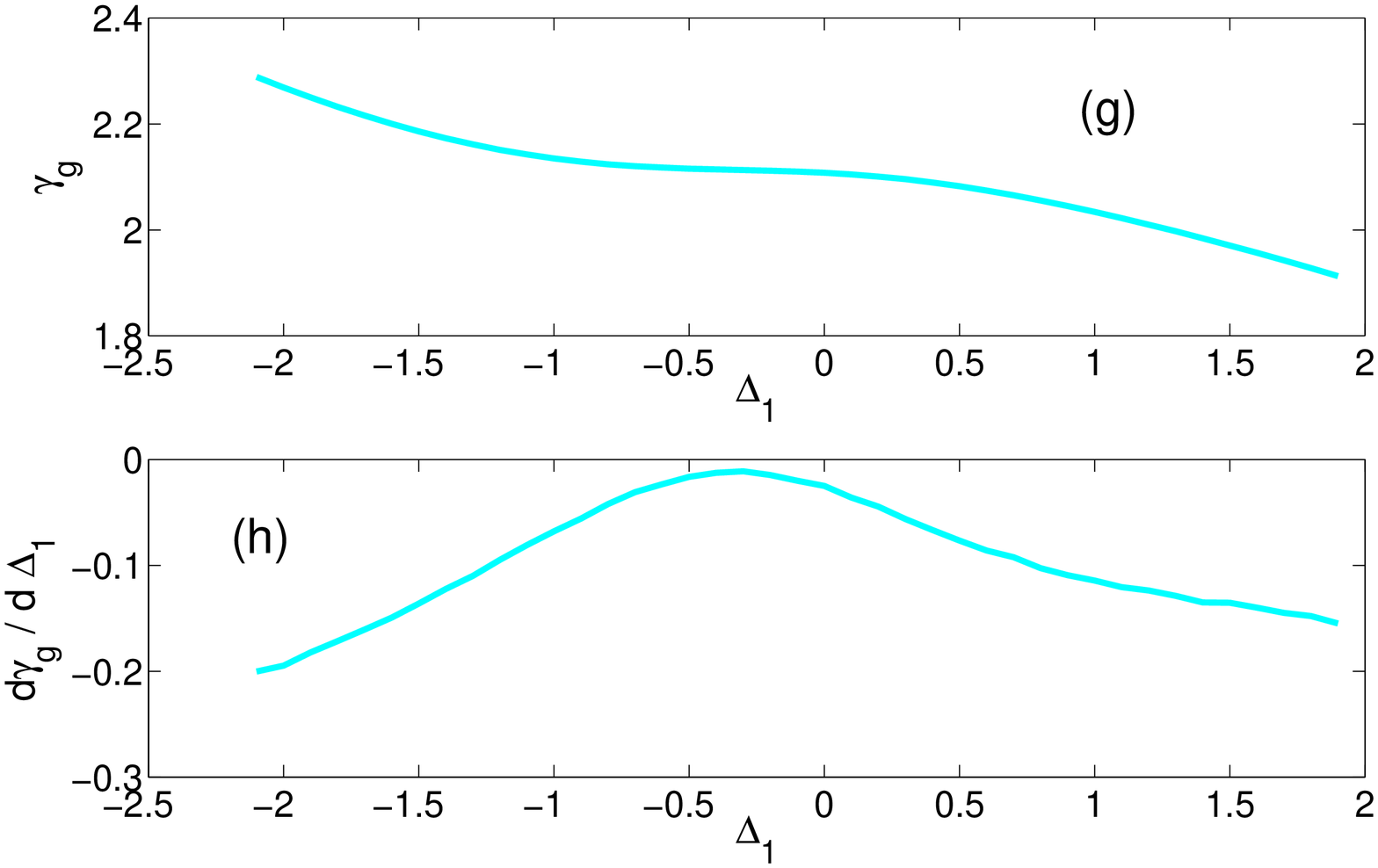} 
\includegraphics[width=6.5cm]{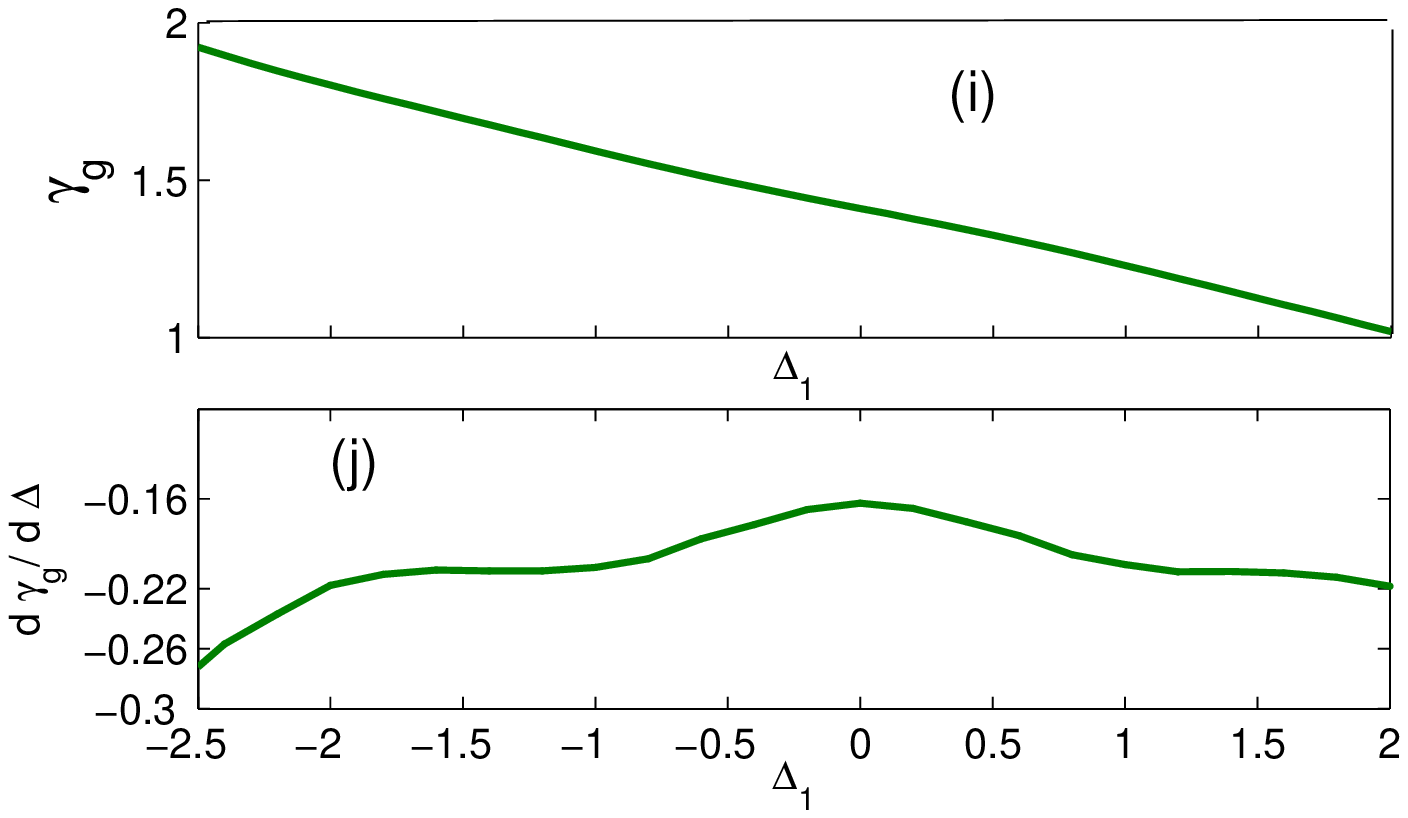}
\caption{Variation of the geometric phase and its derivative with the detuning parameter $\Delta_1$. Parameter values of a) and b) are $\Omega_1=\Omega_2=6, \Delta_2=0$  c) and d) $\Omega_1=3,\Omega_2=6, \Delta_2=0$,  e) and f) $\Omega_1=6, \Omega_2=3, \Delta_2=0$. g) and h) $\Omega_1=\Omega_2=6,\Delta_2=3$, i) and j) $\Omega_1=1.5,\Omega_2=6,\Delta_2=0$}
\label{fig:5}
\end{figure}

We  evaluate the three eigenvalues and eigenvectors numerically up to all orders in the two photon detuning $\Delta$ and determine the geometric phase as a weighted sum of the individual phase corresponding to each eigenstate. Since we are addressing  hyperfine transitions the eigenvalues are always non-degenerate in the parameter regimes of interest, in other words the hyperfine separations are larger compared to the Rabi frequency of the applied fields. 
\begin{figure}
\centering
\includegraphics[width=8.0cm]{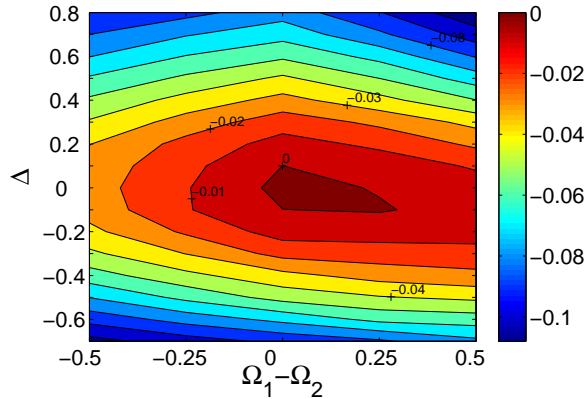} 
\caption{Stability of $\gamma_g$ with respect to fluctuations in $\Delta$ and $\Omega_1-\Omega_2$. }
\label{fig:4}
\end{figure}

All the system parameters are rendered dimensionless by scaling with the atomic lifetime $\gamma \approx 1MHz$.  We restrict our study to the parameter regime $\Delta, \Omega \le 6.0 $ which is the lifetime of the first excited state. 
We would like to mention here that  in the numerical program,  $\gamma_g$ has been initialized to zero for minimum values of the parameter $\Delta_1$. 

In the previous section, for the scheme I, we have seen that the state is dominantly  entangled when
$\Omega_1\approx \Omega_2$ and  separable otherwise. Even in the case of the real system (scheme I), we may  identify three regions in the parameter space: (i) when $\Omega_1<< \Omega_2, \Delta_{1,2}=0$ the two photon state is dominantly in a pure state with very weak entanglement , (ii)  when $\Omega_1\approx \Omega_2, \Delta_{1,2}=0$ the state is almost pure  Bell state of the type $[|00 \rangle - |11\rangle]/\sqrt{2}$  and (iii)  when $\Omega_1>> \Omega_2, \Delta_{1,2}=0$ the state is mixed with weak entanglement. Non-zero detuning ($\Delta \neq 0$) further affects both purity and entanglement.  A distribution of concurrence as a function of the two photon detuning $\Delta$ and $\Omega_1-\Omega_2$ is shown in Fig.3b for the real system with decoherence. The state seems to be show maximum entanglement near $\Omega_1 \approx \Omega_2$ and shows weaker entanglement for $\Omega_1 <<\Omega_2$ and $\Omega_1 >> \Omega_2$. It is clear that even in the presence of decoherence the qualitative features of entanglement distribution is similar even though there are quantitative changes. This is not very surprising  given the fact that there is no unique entanglement measure for mixed states.

We illustrate in detail the variation of the geometric phase  in the neighborhood of $\Delta_1=0$ with $\Delta_2=0$. This is effectively varying the two photon detuning $\Delta$.  Figs. 5. a, b show the variation of $\gamma_g$ and the rate of change of $\gamma_g$, respectively with $\Delta_1$  near $\Delta_1=0$ for $\Omega_1=\Omega_2=6 $. The main feature we observe is that  the geometric phase varies slowly near the region of entanglement which we showed to be true in the case of the ideal system earlier (Fig.3b). 
 Observe that $\gamma_g$ is constant in the region $-0.25< \Delta_1< 0.25$ (Fig.5 a,b). This is substantiated by the vanishing of the derivative in this region thus indicating that the geometric phase in this case is stable under  small perturbations in the neighborhood of $\Delta_1=0$.  The numerical values  reveal\cite{sns_vr} that the population is almost equal in the states $ |00\rangle$ and $|11\rangle$ with non-zero correlation which is typical of a Bell like state. Again in Figs. 5 g, h which corresponds to the parameter values $\Omega_1=\Omega_2=6$ and $\Delta_2=3$, the geometric phase  $\gamma_g$ does not change much and the derivative is nearly zero in the region $-0.5< \Delta_1 < 0.$ The  entanglement distribution in these two cases  is nearly the same with the two photon state being an almost pure Bell state. We have verified that the variation of the density matrix elements in these two cases is less than $2\%$.  
 In the  case of Figs. 5 (c), (d), the parameter values are $\Omega_1=3, \Omega_2=6,\Delta_2=0$.  Observe that rate of variation of $\gamma_g$ is more rapid compared to (b) and (h) indicating an overall weaker entanglement.
 The slope decreases to a minimum at $\Delta_1=0$, implying that the state is maximally entangled here. However, $\gamma_g$ is never constant in this regime which means it is not stable under perturbation in the parameter $\Delta$. A look at the density matrix elements reveals \cite{sns_vr} that the state is an almost pure state but with  weak entanglement.

Lastly,   we  address the question whether geometric phase can distinguish between  pure separable state and mixed state.
As shown if Fig.3b, the distribution of concurrence for the real system does not seem to distinguish between the pure separable state and the mixed state. In fact this is true of other measures like negativity and fidelity. However, the variation of geometric phase seems to show this distinction.  For the parameter values $\Omega_1=6, \Omega_2=3$ the state is a mixed state \cite{sns_vr} and the variation is illustrated in Fig.5 e,f.  Here $\gamma_g$ is never a constant and the slope which corresponds to the rate of change of $\gamma_g$ shows sharp fluctuations in the neighborhood of $\Delta_1=0$. While for a pure separable state $\Omega_1=1.5, \Omega_2=6$, the derivative  shows a smooth variation as illustrated in Fig.5 i, j. 
Thus geometric phase is unique in the sense that it carries information on both purity/mixedness and entanglement.

  In Fig. 6 we present the details of this plateau where $\gamma_g$ is stable. The plateau
 in the center marked zero corresponds to the  region where $\gamma_g$ does not change. This is sorrounded by region where $\gamma_g$ changes by $1\%$. The negative sign indicates the decrease in $\gamma_g$. The figure illustrates a total decrease of $\gamma_g$ by $10\%$. Thus, in a given interval of the control parameter, faster the sweep of $\gamma_g$ (with respect to the control parameter),  weaker the entanglement and vice versa. This is consistent with the distribution of concurrence shown in Fig.3b.

\section{Summary and Conclusions}
We have studied the evolution of the geometric phase in the  general set up namely non-adiabatic, non-cyclic and non-unitary evolution of mixed states. The scheme corresponds to a Hamiltonian system consisting of interacting atom-photon system. We have identified the   control parameters for manipulating the purity/mixedness and entanglement of the emitted two-photon states. We have been able to show the relation between the geometric phase and entanglement in terms of the control parameters analytically for an ideal system without decoherence. For the real system, we show that the variation of the geometric phase is correlated to the variation of the concurrence on the  control parameters  $\Omega_i$ and $\Delta_i,~i=1,2$.  We have shown that $\gamma_g$ is robust under fluctuations of these control parameters  only for pure Bell like states.  We  further show that the variation of the geometric phase is able to distinguish between pure separable and mixed states. These features provide a diagnostic tool for determining the entanglement of the two-photon state.  It is interesting to note that such a connection, between entanglement and geometric phase, has also been found, recently \cite{cs11}, in another system, viz. Heisenberg spin chains undergoing unitary evolution.  
%We hope that it should be possible to measure $\gamma_g$ by a two-photon interferometric set up like the one  recently proposed in \cite{intf}. 

\section{ Acknowledgements} 
One of us, SNS thanks the Department of Science and Technology, India, for financial assistance through the WOS-A scheme. We would also like to thank  Joseph Samuel and V. Ravishankar for useful discussions.

\bibliographystyle{apsrev}

\begin{thebibliography}{100}
\bibitem{bibitem} This result is the quantum analog of the fact that the amplitude of the radiation emitted by an oscillating dipole is proportional to the the dipole operator at retarded time.
\end{thebibliography}

\end{document}